\documentclass[twocolumn,runningheads,natbib]{svjour2}
\journalname{Astrophysics and Space Science}
\smartqed  
\usepackage{graphicx}
\newcommand{\rmmat}[1]{{\hbox{\rm #1}}}
\newcommand{\rmscr}[1]{{\rmmat{\scriptsize #1}}}
\newcommand{\be}{\begin{equation}}
\newcommand{\ee}{\end{equation}}

\newcommand{\gtrsim}{\stackrel{>}{\sim}}
\newcommand{\plotone}[1]{\centering \includegraphics[width=8cm]{#1}}

\begin{document}

\title{QED can explain the non-thermal emission from SGRs and AXPs : Variability
\thanks{The Natural Sciences and Engineering Research Council of Canada,
Canadian Foundation for Innovation and the British Columbia Knowledge
Development Fund supported this work. J.S.H. is a Canada Research Chair.
}
}


\author{Jeremy S. Heyl}


\institute{Jeremy S Heyl \at
              University of British Columbia
              6224 Agricultural Road, Vancouver, BC, Canada V6T 1Z1 
              \email{heyl@phas.ubc.ca}
}

\date{Received: date / Accepted: date}

\maketitle

\begin{abstract}
  Owing to effects arising from quantum electrodynamics (QED),
  magnetohydrodynamical fast modes of sufficient strength will break
  down to form electron-positron pairs while traversing the
  magnetospheres of strongly magnetised neutron stars. The bulk of the
  energy of the fast mode fuels the development of an
  electron-positron fireball. However, a small, but potentially
  observable, fraction of the energy ($\sim 10^{33}$ ergs) can
  generate a non-thermal distribution of electrons and positrons far
  from the star. This paper examines the cooling and radiative output
  of these particles. Small-scale waves may produce only the
  non-thermal emission. The properties of this non-thermal emission in
  the absence of a fireball match those of the quiescent, non-thermal
  radiation recently observed non-thermal emission from several
  anomalous X-ray pulsars and soft-gamma repeaters.  Initial estimates
  of the emission as a function of angle indicate that the non-thermal
  emission should be beamed and therefore one would expect this
  emission to be pulsed as well.  According to this model the
  pulsation of the non-thermal emission should be between 90 and 180
  degrees out of phase from the thermal emission from the stellar
  surface.
\end{abstract}

\section{Introduction}
\label{sec:introduction}

Simply put magnetars are neutron stars whose magnetic fields dominate
their emission, evolution and manifestations.  In the late 1970s and
early 1980s,  a fleet of sensitive detectors of high-energy radiation
uncovered two new phenomena, the soft-gamma repeater and the anomalous
x-ray pulsar.  Strongly magnetized neutron stars provide the most
compelling model for both types of object, and observations over the
past few years indicate that these phenomena are two manifestations of
the same type of object.   Soft-gamma repeaters exhibit quiescent
emission similar to that of anomalous x-ray pulsars
\cite[e.g][]{1994Natur.368..432R,1994Natur.368..127M,1996ApJ...463L..13H,1999ApJ...510L.111H},
and anomalous x-ray pulsars sometimes burst
\cite[][]{2002Natur.419..142G,2003ApJ...588L..93K}.
What makes magnetars a hot topic of
research is the rich variety of physical phenomena that strong
magnetic fields exhibit.

This article will focus on the quiescent emission from these
interesting objects rather than the bursts themselves (The reader may
wish to refer to the seminal work of \cite{Thom95} for details of the
burst but may also want to look at \cite{Heyl03sgr} for an
alternative).  Furthermore, the article will concentrate on a possible
model for the recently detected non-thermal emission from these
objects.

In earlier work, Lars Hernquist and I analysed wave propagation
through fields exceeding the quantum critical value $B_\rmscr{QED}
\equiv m^2c^3/e\hbar \approx 4.4\times 10^{13}$ G, and demonstrated
circumstances under which electromagnetic \citep{Heyl98shocks} and
some MHD waves, particularly fast modes \citep{Heyl98mhd} evolve in a
non-linear manner and eventually exhibit discontinuities similar to
hydrodynamic shocks, owing to vacuum polarisation from quantum
electrodynamics (QED).  In Paper I \cite[][]{Heyl03sgr}, we developed
a theory to account for bursts from SGRs and AXPs based on ``fast-mode
breakdown,'' in which the wave energy is dissipated into electron-positron
pairs when the scale of these discontinuities becomes comparable to an
electron Compton wavelength.  We showed that, under appropriate
conditions, an extended, optically thick pair-plasma fireball would
result, radiating primarily in hard X-rays and soft $\gamma$-rays.  In
Paper II \cite[][]{Heyl05sgr} we developed the theory of the
non-thermal emission from the fast-mode cascade under a series of
assumptions and approximations.  In particular, MHD fast modes of
insufficient amplitude to generate an optically thick fireball will
still dissipate through pair-production, seeding non-thermal emission.
\cite{Heyl03sgr} estimates the Thomson optical depth through the pair
plasma created by a passing fast mode (their Fig.~3).  If this optical depth
exceeds unity, one would expect a fireball to form.

In what follows, I extend our previous investigation of fast-mode
breakdown to estimate the spectrum of non-thermal emission expected
outside the region containing an optically thick fireball with
particular emphasis on the assumptions made in Paper II and on the
angular dependence of this emission.

\section{The Observed Non-Thermal Emission}
\label{sec:observed-non-thermal}

What would be the typical flux of these small-scale fast modes?
\cite{Thom96} and \cite{Heyl98decay} have argued that the quiescent emission of SGR and
AXP neutron stars may be powered by the decay of the magnetic field.
The quiescent thermal emission may only be a small fraction of the
total energy released by the decay of the magnetic field.  Recent
observations of SGRs and AXPs indicate that the thermal radiation may
indeed be just the tip of the iceberg
\cite[][]{2004astro.ph.11696M,2004astro.ph.11695M,2004ApJ...613.1173K};
therefore, following the discussion of Paper II, the amplitude of the
thermal and non-thermal radiation are taken to be comparable, and the
pair cascade operates beyond a certain radius from the star or
equivalently below a certain magnetic field strength
($B_\rmscr{max}$).  This model has two parameters, $B_\rmscr{max}$
determines the position of the two breaks in the spectrum and the
total normalisation.  The process of fast-mode breakdown according to
Paper II predicts a particular relationship between the location of
the two breaks in the spectrum and particular slopes between and
beyond the breaks.

\begin{figure}
\plotone{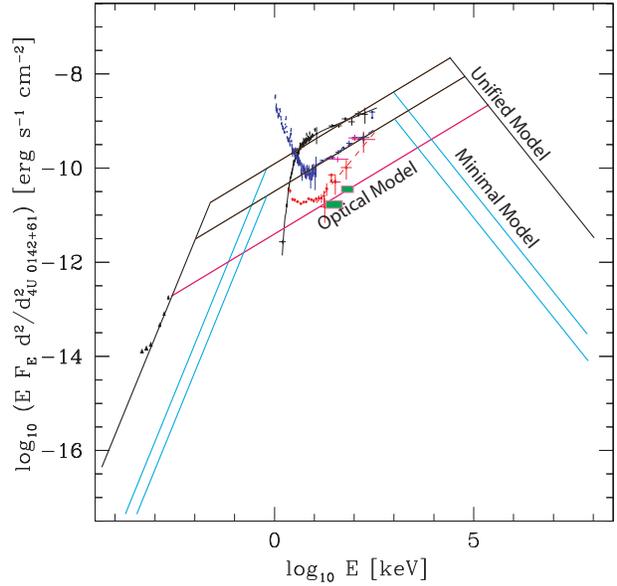}
\caption{The spectrum produced by fast-mode breakdown is superimposed
over the observed thermal and non-thermal emission from several AXPs
and SGRs for models that fit either the optical or INTEGRAL data
solely and one that fits both sets of data.  The unabsorbed optical
data are from \cite{2000Natur.408..689H} via
\cite{2004astro.ph..4144O} for AXP 4U~0142+61.  The uppermost black
symbols are the hard X-ray band are from \cite{2004astro.ph.11696M}
for SGR~1806-20.  \cite{2004astro.ph.11695M} obtained similar results
for the SGR.  The middle sets of points in the hard X-ray data (blue
is total flux and red is pulsed flux) are from
\cite{2004ApJ...613.1173K} for AXP 1E~1841-045.  The green squares
plot the INTEGRAL data reported by \cite{2004ATel..293....1D} for
AXP~4U~0142+61. The \cite{2004ATel..293....1D} results
are normalised using the observations of the Crab by \cite{1989ApJ...338..972J}.  We
scaled the emission from the three sources by assuming that they all
lie at the distance of AXP~4U~0142+61.  The assumed distances are 3 kpc for
AXP~4U~0142+61 \cite[][]{2000Natur.408..689H}, 7.5 kpc for AXP
1E~1841-045 \cite[][]{1992AJ....104.2189S} and 15 kpc for SGR~1806-20
\cite[][]{2004astro.ph.11696M}.}
\label{fig:allspec}
\end{figure}

Figure~\ref{fig:allspec} shows the observed broad-band spectrum of
several AXPs and SGRs.  To compare the spectra the more distant
objects (AXP~1E~1841-045 and SGR~1806-20) whose hard X-ray emission
was discovered with INTEGRAL have been placed at the distance of
4U~0142+61 whose optical emission \cite[][]{2000Natur.408..689H} is
very likely to be nonthermal \cite[][]{2004astro.ph..4144O}.  The
observed spectra depart from a power law for $E<1~$keV because in this
region the thermal radiation from the neutron star begins to dominate
the non-thermal component.

Because the location of the two breaks in the spectrum both depend on
the strength of the magnetic field at the inner edge of the breakdown
region, the presence of extensive nonthermal optical emission
indicates that the non-thermal hard X-ray emission should peak at
about 30~MeV, a factor of two hundred beyond the observed spectrum.
The best limits in this energy range are provided by Comptel
\cite[][]{2006ApJ...645..556K} and appear to exclude the ``optical model'' for
4U~0142+61.

This conclusion assumes that the sources SGR~1806-20 and
AXP~1E~1841-045 have a similar optical excess to 4U~0142+61.  A more
conservative assumption would be that the hard X-ray emission does not
extend far beyond the observations from INTEGRAL with spectral breaks
at about 1~MeV and 650~eV.  This situation is somewhat natural.  The
fast-mode cascade is limited to pairs with sufficient energy to
produce photons with $E > 1$~MeV that can subsequently pair produce.
Lower energy electrons simply cool, giving the observed cooling
spectrum in the hard X-rays.  The total energy in the non-thermal
emission is also reduced by a factor of a few.  In the context of the
fast-mode cascade it is difficult to have $E_\rmscr{break} < 2 m c^2$.
This model is denoted as the ``Minimal Model'' in
Fig.~\ref{fig:allspec} because $E_\rmscr{break}$ takes on the minimal
value that makes sense physically; i.e $\approx 1$~MeV; this is
equivalent to assuming that magnetic field is much weaker than the
quantum-critical limit in the pair-production region.

The analysis reviewed here from Paper I assumed that the fast-mode was
travelling perpendicular to the field and that the amplitude of the wave
and the magnetic field where the shock forms are independent.  The
more detailed analysis that follows shows that the location of shock
formation depends on the product of initial amplitude of the wave and
its wavenumber, its direction of propagation and the strength of the
dipole component of the star's magnetic field.  The latter two
parameters could potentially be determined independently of the
emission spectrum, leaving a single free parameter to describe the
emission, the product of initial amplitude of the wave and
its wavenumber.

\section{Spectrum}
\label{sec:spectrum}

The previous calculation of the spectrum assumed that the energy
dissipated by the wave over a given range of magnetic field strengths
is constant and that the dissipation only occurs below a particular
magnetic field strength.  A direct calculation the evolution of the
wave as it travels away from the star relaxes both of these
assumptions.  This determines that amount of energy dumped into pairs
at various magnetic field strengths; however, it does not determine
how this energy is finally dissipated.  It can be emitted locally and
promptly as classical synchrotron radiation or Landau transition
radiation or elsewhere as curvature emission.

\subsection{Weak-Field Regime}
\label{sec:weak-field-regime}

If the magnetic field is much weaker than the quantum-critical limit,
the spectrum of the cooling pairs at a particular value of the
magnetic field is given by
\begin{equation}
\frac{d E}{d E_\gamma} = \frac{E_\perp}{2(E_\rmscr{break}^{1/2} - E_0^{1/2})} E_\gamma^{-1/2}
\label{eq:1}
\end{equation}
where $E_0=m_e c^2 \xi$, $E_\rmscr{break} \approx m_e c^2 / (40 \xi)$
and $\xi=B/B_\rmscr{QED}$ for $B \ll B_{QED}$.  This assumes that the
final generation of pairs emits classical synchrotron radiation as
described in Paper II and that only the momentum perpendicular to the
magnetic field is dissipated as synchrotron emission.

\subsection{Strong-field Regime}
\label{sec:strong-field-regime}

For $\xi \gtrsim 0.1$ the analysis of the preceding section breaks
down.  This is signaled by the fact that $E_0 \gtrsim
E_\rmscr{break}$, so Eq.~(\ref{eq:1}), for example, does not make
sense.  In this regime both the processes of pair production and
synchrotron emission are strong affected.  Additionally photon
splitting may play a role.  Restricting the results to $\xi \ll 1$
yields the results of \cite{Heyl05sgr}.

Specifically in the strong-field regime, neither the primary pairs nor
the secondaries can be assumed to have relativistic motion
perpendicular to the magnetic field.  The shock travels at a velocity
\cite[][]{Heyl97index},
\begin{equation}
v = \frac{c}{n_\perp} \approx c \left [ 1 - \frac{\alpha_\rmscr{QED}}{4\pi} X_1\left (\frac{1}{\xi} \right ) \sin^2 \theta \right ]^{-1}
\label{eq:2}
\end{equation}
where the function is defined in \cite{Heyl97hesplit} and well approximated by 
the fitting formula 
\begin{equation}
X_1 \left (\frac{1}{\xi} \right ) \approx -\frac{14}{45} \xi^2 \frac{1+\frac{6}{5}\xi}{1+\frac{4}{3} \xi + \frac{14}{25}\xi^2}
\end{equation}
from \cite{2004ApJ...612.1034P}.

The primary pairs typically have a Lorentz factor of
\begin{equation}
\gamma = \left (1 - \frac{v^2}{c^2} \right )^{-1/2} \approx \frac{1}{|\xi \sin\theta|} \left ( \frac{45\pi}{7\alpha_\rmscr{QED}} \frac{1+\frac{4}{3}\xi+\frac{14}{25}\xi^2}{1+\frac{6}{5}\xi} \right )^{1/2}.
\label{eq:3}
\end{equation}
The typical Landau level occupied by these primary pairs is
\begin{equation}
n \approx \frac{45}{14} \frac{\pi}{\alpha_\rmscr{QED}} \frac{1}{\xi^3} \frac{1+\frac{4}{3} \xi + \frac{14}{25} \xi^2}{1 + \frac{6}{5} \xi}
\end{equation}
independent of angle, so for $\xi>27$ only the ground Landau level is
occupied by the primary pairs.  \cite{1995AuJPh..48..571U} found a
limit of $\xi>0.1$ for pairs to be created of pairs in the ground
Landau level.  The limit found here is quite a bit different simply
because the pair cascade discussed here is due to photons that are
essentially created above the pair-production threshold in the shock.
The direction of the photons is constant relative to the field
direction; it is their energy that increases above the pair-production
threshold; they carry much momentum perpendicular to the field.  On
the other hand, \cite{1995AuJPh..48..571U} examine the pair production
by photons travelling nearly parallel to the magnetic field.  The
momentum perpendicular to the field is small so the ground Landau
level is preferred at much lower fields.

The Landau level of the secondaries is a factor of sixteen smaller, so
for $\xi>8$ only the ground level is occupied by the secondary pairs.
In these cases there is no prompt emission at all from the pairs.  The
energy of the initial photon goes into the rest mass of the pairs and
their motion along the magnetic field.  If $n$ is not large, the
emission is better described by cyclotron emission than by
synchrotron.  The emission is not cutoff by considerations of the
total pair-production optical depth as in the weak-field case
\cite[][]{Heyl05sgr} but by the Landau levels of the pairs produced in
the cascade.  In the weak-field limit, \cite{Heyl05sgr} found that the
energy of the final pairs in the cascade was a factor of 1,700 lower
than the primaries, so for $\xi>0.08$ these final pairs occupy the
ground Landau level and the synchrotron analysis of the cascade
emission must be modified; the limiting field for the final pairs in a
cascade to be in ground Landau level is similar to that found by
\cite{1995AuJPh..48..571U} for the pulsar cascade.  Comparing the
values of $E_0$ and $E_\rmscr{break}$ yields the same limiting field.

\subsection{The Total Spectrum}
\label{sec:total-spectrum}

The analysis of the pair cascade in the strong-field limit is beyond
the scope of this small contribution.  Instead I shall focus on the
regime where the entire cascade is in the weak-field limit.  This may
occur for a very strong wave propagating near a weakly magnetised star
or for a relatively weak wave travelling near a strongly magnetised
star.

In an arbitrary magnetic field, the evolution of a fast mode can be
parametrized by the integral of the ``opacity to shocking''
\begin{equation}
\kappa = \frac{k \sin\theta}{2} \frac{b}{B} \frac{e^2}{hc}
\left [ \sin^2 \theta \kappa_1
+ \left ( 4 - 3 \sin^2 \theta \right ) \kappa_2
\right ]
\label{eq:kappa}
\end{equation}
along the path of the wave.  
In this expression, $b$ is the amplitude of the
magnetic field of the fast mode, $B$ is the amplitude of the local
magnetic field, $\theta$ is the angle between the direction of
propagation and the magnetic field and $k$ is the wavenumber of the
fast mode.

Although this quantity is not strictly an opacity in the sense of
radiative transfer, it does characterize the evolution of the wave in
an analogous way.  In particular, the shape of the wave depends on the
optical depth for shocking
\begin{equation}
\tau = \int \kappa d l.
\end{equation}
The shock forms at $\tau=1$, and the power carried by the wave
decreases as $\tau$ increases above unity.

\begin{figure}
\plotone{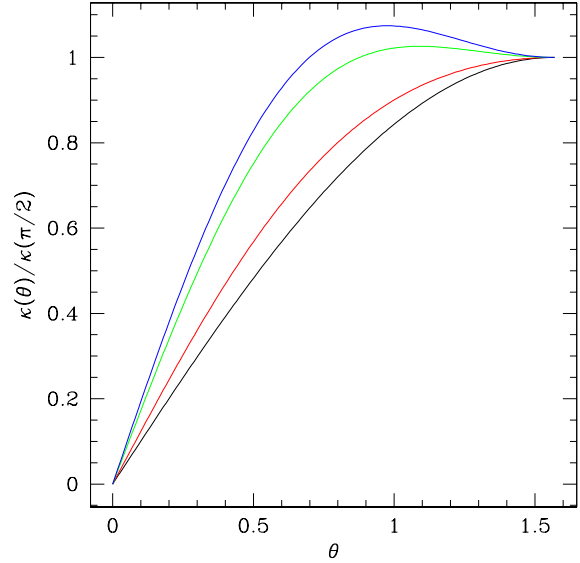}
\caption{The opacity as a function of angle at various magnetic field
  strengths.  The curves from bottom to top are for $\xi=0.1,1,10$ and
  100.  For weak fields, the opacity peaks at $\theta=\pi/2$.  For
  strong fields the opacity peaks at $\theta=\cos^{-1}\left
    (1/\sqrt{3}\right ) \approx 55^\circ$.  }
\label{fig:kappang}
\end{figure}

\begin{figure}
\plotone{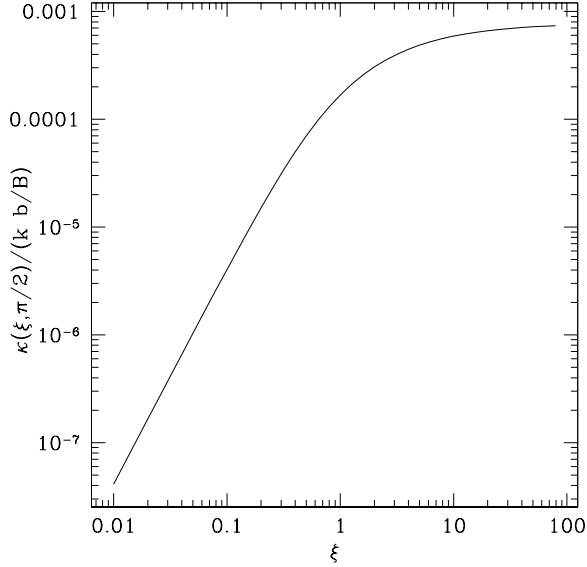}
\caption{ The opacity as a function of the strength of the field at
  $\theta=\pi/2$ in units of $k$ with $b=B$.  The reciprocal of the
  ordinate is the number of wavelengths that a wave of amplitude equal
  to the background field will travel before forming a shock. }
\label{fig:kappaxi}
\end{figure}

The quantity in brackets is the generalization of the function $\xi
F(\xi)$ where $F(\xi)$ was defined by \cite{Heyl98shocks}. The two
terms are
\begin{eqnarray}
\kappa_1 &=&  -\xi^{-3} X_0^{(3)} \left ( \frac{1}{\xi} \right ) \\
\kappa_2 &=& X^{(2)}_0 \left (\frac{1}{\xi} \right ) \xi^{-2} -  X_0^{(1)} \left ( \frac{1}{\xi} \right )\xi^{-1}  
\end{eqnarray}
Fig.~\ref{fig:kappang} and Fig.~\ref{fig:kappaxi} depict the opacity
as a function of angle and magnetic field strength.  For weak fields
the opacity is proportional to $\sin\theta$; while in strong fields,
it is proportional to $\sin\theta \left (2 - \sin^2\theta\right)$.

The definition of $X_0(x)$ from \cite{Heyl97hesplit} yields
\begin{eqnarray}
\kappa_1 &=& 
\frac{2}{3} + \frac{1}{\xi} + \frac{1}{\xi^2} - \frac{1}{2\xi^3} {\rm \Psi}^{(1)} \left ( \frac{1}{2\xi} \right ) \\
&\approx& \frac{8}{15} \xi^2 - \frac{32}{21} \xi^4 + \frac{128}{15} \xi^6 - \frac{2560}{33} \xi^8 +  {\cal O} \left ( \xi^{10} \right ) \\
&\approx& \frac{2}{3} - \frac{1}{\xi} + \frac{1}{\xi^2} - \frac{\pi^2}{12} \frac{1}{\xi^3} + \frac{\zeta(3)}{2} \frac{1}{\xi^4} + {\cal O} \left ( \frac{1}{\xi^5} \right ) \\
\kappa_2 &=& 
 \frac{2}{3} + \frac{\ln \left ( 4\pi\xi \right ) - 2 \ln {\rm \Gamma} \left (\frac{1}{2\xi} \right )+1}{\xi} 
 +  \frac{{\rm \Psi} \left ( \frac{1}{2\xi} \right )-1}{\xi^2} \\
&\approx& \frac{8}{45} \xi^2 - \frac{32}{105} \xi^4 + \frac{128}{105} \xi^6 - \frac{2560}{297} \xi^8 +  {\cal O} \left ( \xi^{10} \right ) \\
&\approx& \frac{2}{3} + \frac{\ln \left ( \frac{\pi}{\xi} \right ) }{\xi} - \frac{1}{\xi^2} + \frac{\pi^2}{24} \frac{1}{\xi^3} - \frac{\zeta(3)}{6} \frac{1}{\xi^4} + {\cal O} \left ( \frac{1}{\xi^5} \right )
\end{eqnarray}
where ${\rm \Psi} (x) \equiv d\ln {\rm \Gamma} / dx$, ${\rm
  \Psi}^{(1)}(x) = d{\rm \Psi}/dx$ and $\zeta(3) \approx 1.202$.  Both
the opacity and the energy available for prompt synchrotron emission
drop as the angle of propagation departs from perpendicular.

The total optical depth is simply the integral of the opacity over the
path of the wave.  If the wave is spherical, the value of $b$
decreases as $1/r$ as the wave travels away from the star.  For
simplicity the angle of the wave propagation with respect to the field
is held constant; this is appropriate for waves travelling radially
and holds approximately for waves that have travelled several stellar
radii from the surface.   The dissipation of the energy of the wave is
given by Eq.~72 in \cite{Heyl98shocks} and Fig.~2 in \cite{Heyl03sgr}.

Summing over all of the magnetic fields where the dissipation occurs yields
\begin{equation}
\frac{d E_\perp}{dB} = -(1-\cos\theta) \frac{d P}{d B} 
\label{eq:4}
\end{equation}
where $P$ is the power remaining in the wave and $\theta$ is the angle
that the direction of propagation of the wave makes with the magnetic
field.  The angular dependence here comes from the assumption that
only the component of the momentum perpendicular to the field yields
prompt and local emission.  Furthermore, this equation also assumes
that the emitting pairs are produced in a weak-field region; they
occupy high Landau levels, so the classical treatment of synchrotron
emission holds.  

\begin{figure}
\plotone{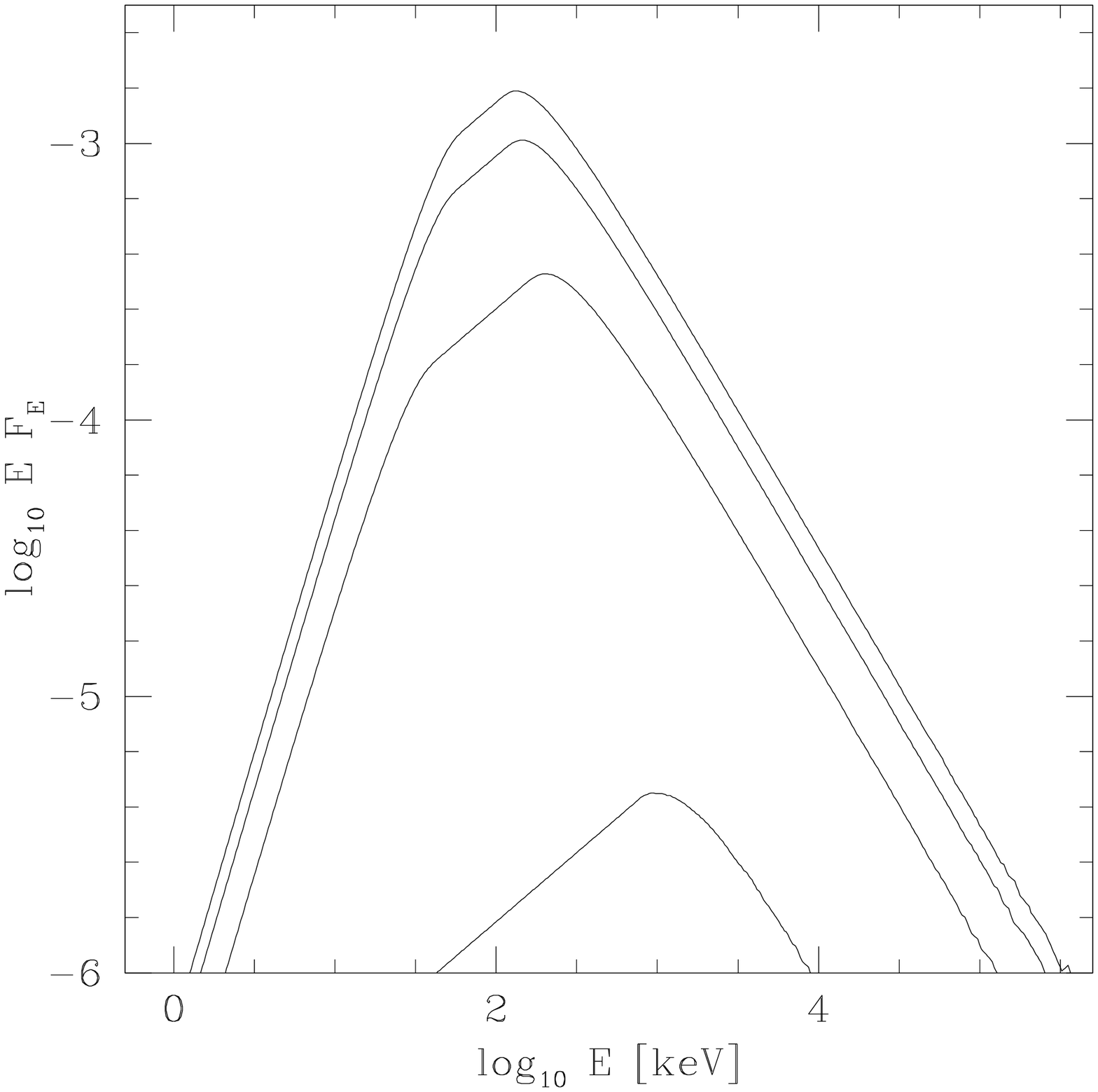}
\caption{Total spectrum for strong-field neutron star
  ($B_\rmscr{NS}=30 B_\rmscr{QED}$) as a function of angle.  The
  initial wave has $\lambda \sim 100$~m and $b \sim 2 B_\rmscr{QED}$.
  The curves from shortest to tallest are for
  $\theta=60^\circ,70^\circ$,$80^\circ$ and $90^\circ$.}
\label{fig:strong}
\end{figure}

For large distances from the neutron star, $dP/dB$ is a
constant and the fields are weak; this in combination with the results
of \S~\ref{sec:weak-field-regime} yields the spectra given in
\cite{Heyl05sgr} and shown in Fig.~\ref{fig:strong} where the opacity 
depends on angle as in Eq.~(\ref{eq:kappa}).  The vertical
axis gives the fraction of the energy of the initial wave that is
dissipated over a factor of $e\approx 2.7$ in energy. For the weak-field
treatment to be accurate, both the frequency and the amplitude of the
wave must be sufficiently small to delay the formation of the shock until
the weak-field regime.  This results in a wave that only dissipates a
small fraction of its energy into pairs.

In this case the total optical depth to shocking is only slightly
greater than unity ($\tau_{90}=1.08$) so the spectrum depends strongly
on the angle of propagation as depicted in Fig.~\ref{fig:strong}.  The
radiation is emitted only if $\tau>1$, so the radiation is emitted
only for $\sin \theta \tau_{90}> 1$ if the bulk of the opacity was in
weak magnetic fields, and $\sin\theta (2 - \sin^2\theta) \tau_{90} >
1$ if it was in weak magnetic fields.  Because the star in this case
has $\xi=30$ but the shock does not form until $\xi<0.1$, one would
expect the cutoff to lie between these extremes.  These two estimates
yield cutoff angles of $67^\circ$ and $33^\circ$.  The detailed
calculations find that the emission cuts off within about $59^\circ$
of the field direction.

\begin{figure}
\plotone{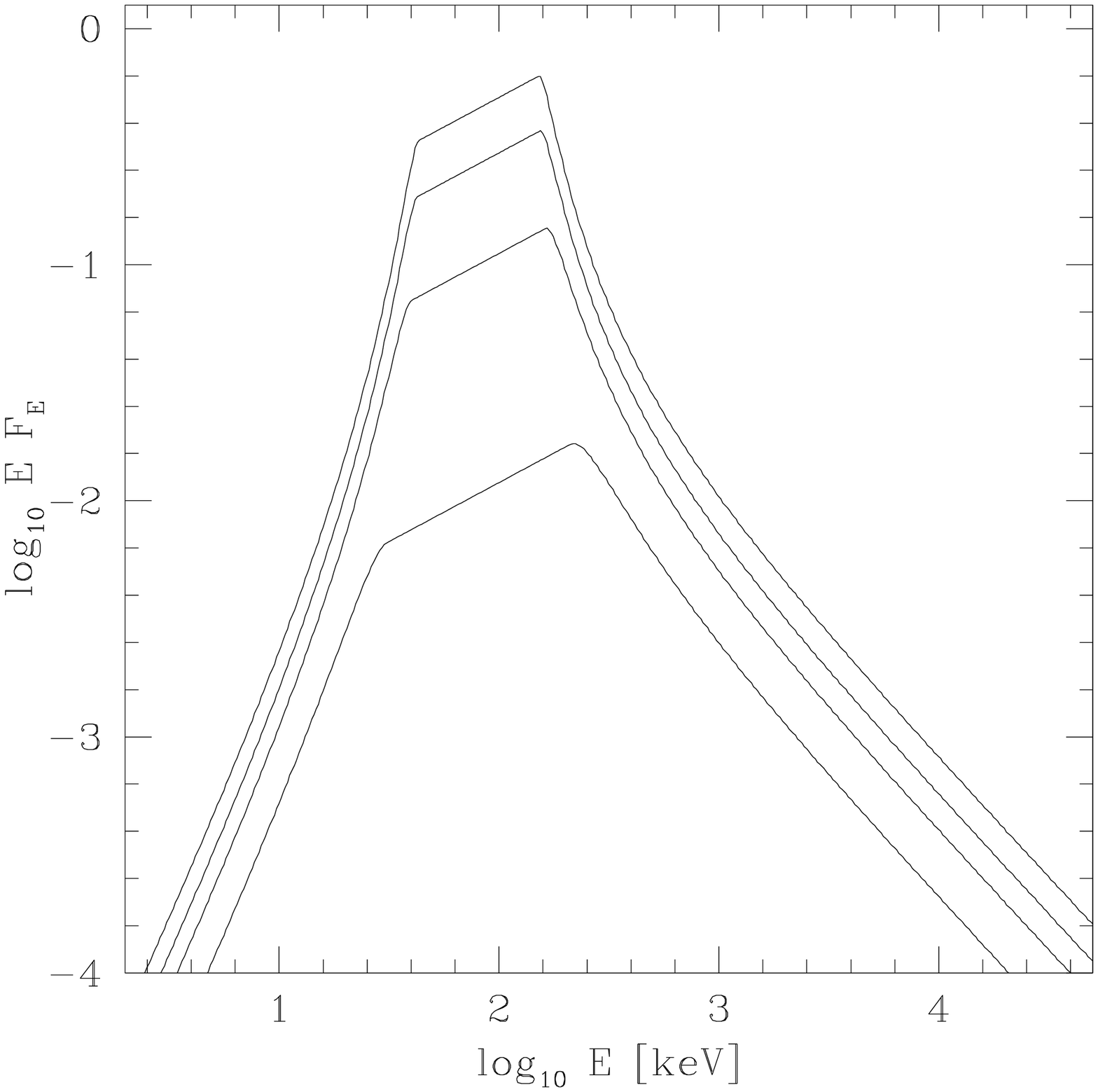}
\caption{Total spectrum for weak-field neutron star ($B_\rmscr{NS}=0.1
  B_\rmscr{QED}$) as a function of angle.  The initial wave has
  $\lambda \sim 20 \mu$ and $b \sim 0.01 B_\rmscr{QED}$.  The curves from shortest to tallest are for
  $\theta=22.5^\circ,45^\circ,67.5^\circ$ and $90^\circ$.}
\label{fig:weak}
\end{figure}

If the field at the surface of the star is less than about one-tenth
of $B_\rmscr{QED}$, the discussion of \S~\ref{sec:weak-field-regime}
apply, regardless of where the shock forms, yielding the results of
Fig.~\ref{fig:weak}.
 
In this case the shock may form relatively close to the star and a
large fraction of the initial energy in the wave is dissipated over a
narrow range of magnetic field strengths where the shock forms.  The
bulk of the energy may be released in the regime where $dP/dB$ is not
constant so the characteristic spectrum discussed in \cite{Heyl05sgr}
does not emerge.  For large angles with respect to the magnetic field,
the shock forms further from the star and the bulk of the energy
dissipates in the constant-$dP/dB$ regime.  Again the emission cuts
off beyond a critical angle.  Here $\tau_{90}=6.82522$ yielding a
cutoff angle of about 8$^\circ$.  Because the optical depth in the
perpendicular direction is so large, the bulk of the angular
dependence in the emission comes not from the opacity but from the
$(1-\cos\theta)$ factor in Eq.~\ref{eq:4}.

\section{Comparison with Observations}
\label{sec:comp-with-observ}

The model outlined in the preceding sections yields at least one
important prediction.  The hard x-ray emission if it is produced by
fast-mode breakdown will be largest where the line of sight to the
pulsar is perpendicular to the local magnetic field.  This happens
when the equatorial regions of the pulsar face the observer.  On the
other hand, the thermal emission from the pulsar is expected to be
strongest when the polar regions face the observer
\cite[e.g.][]{Heyl97analns}; consequently the hard and soft x-ray
emission should be between 90 and 180 degrees out of phase relative to
each other.

If the magnetic field of the neutron star consists of a
dipole whose dipole moment makes an angle $\beta$ with the spin axis.
Furthermore, the line of sight makes an angle $\alpha$ with the spin
axis.  If $\alpha + \beta \geq 90^\circ$, the hard emission
peaks at a phase
\begin{equation}
\phi = \arccos \left ( -\cot\alpha \cot \beta \right ).
\end{equation}
relative to the main pulse of soft emission (for
$\alpha+\beta>90^\circ$ there can be a second pulse of soft emission
from the second polar region). Because $\alpha$ and $\beta$ lie
between zero and ninety degrees, their cotangents are positive, so
$\phi$ must lie between ninety and one hundred eighty degrees.  On the
other hand if $\alpha+\beta < 90^\circ$, the hard emission peaks
180$^\circ$ out of phase from the soft emission.  

\cite{2006ApJ...645..556K} observed the x-ray emission from several
anomalous x-ray pulsars from 0.5 to 300~keV.  The AXP 1RXS~J1708-4009
exhibits variability similar to that described here.  In the
1.3$-$3.9~keV-band the emission peaks around phase 0.25; while above
8~keV the emission peaks around phase 0.55 (90 degrees away).  The
peak remains at phase 0.55 to the highest energies measured about
300~keV.

The AXP 4U~0142+61 shows a similar effect albeit less dramatically.
At the lower energies that \cite{2006ApJ...645..556K} examined
(0.5$-$1.7~keV), the pulse profile is double peaked with two peaks
about 160$^\circ$ out of phase.  As the energy of the radiation
increases to 50~keV, the second peak increases in amplitude and the
first peak practically vanishes.  The AXP 1E~2259+586 shows similar
trends but the variability is only detectable up to about 25~keV.  On
the other hand the pulse profile of AXP~1E~1841-045 does not change
much from 2.1~keV to 100~keV.  This might mean that this object is an
exception or that the soft thermal emission only begins to dominate at
lower energies.

\section{Discussion}
\label{sec:discussion}

This paper has examined the angular dependence of the non-thermal
emission from AXPs and SGRs in the weak-field regime.  It is a natural
extension of the model for bursts and non-thermal emission presented
in Papers I and II.  When a dislocation of the surface of the neutron
star is sufficiently large, the resulting fast modes will produce
sufficient pairs to make the inner magnetosphere of the neutron star
opaque to x-rays, generating a fireball.  Some small but observable
fraction of the energy initially in the fast modes is dissipated
outside the opaque region yielding a characteristic fast-mode
breakdown spectrum.  A second possibility is that the crust of the
neutron star is constantly shifting over small scales, generating fast
modes whose breakdown is insufficient to produce a fireball.  In this
case we would associate the non-thermal radiation with the quiescent
thermal radiation from the surface of the star.  Both are produced by
the quasi-continuous decay of the star's magnetic field.

Observations of the AXPs and SGRs continue to surprise, as do
theoretical investigations of ultramagnetised neutron stars.  This and
the preceding papers have presented a unified model for the thermal
burst emission and non-thermal emission from ultramagnetised neutron
stars.  The model has few underlying assumptions: magnetars produce
fast modes sufficient to power the non-thermal emission and, more
rarely, the bursts, the magnetic field far from from the star is
approximately dipolar and quantum electrodynamics can account for the
dynamics of pairs and photons in strong magnetic fields.  The model
for the non-thermal emission from a particular fast mode depends only
on the product of the strength of the wave and its wavenumber, the
angle the wave propagates relative to the magnetic field direction and
the strength of the dipole component of the field.  The intensity of
the emission depends strongly on the direction of the initial wave.
Waves that travel sufficiently close to the direction of magnetic
field do not produce non-thermal emission by this mechanism at all.
The emission is therefore beamed perpendicular to the magnetic field
lines, so the non-thermal emission should be between ninety and one hundred
eighty degrees (depending on the viewing geometry) out of phase
relative to the thermal emission which is expected to be strongest
along the magnetic field lines.

Further calculations of the fate of the pairs produced in the strong
field regions ($\xi \gtrsim 0.1$) is needed to understand the spectra
produced by fast-mode breakdown in general; however, the beaming of
the radiation does appear robust.  Further observations can easily
verify or falsify this model and potentially provide direct evidence
for the ultramagnetised neutron stars that power AXPs and SGRs and the
macroscopic manifestations of QED processes that account for their
unique attributes.

\begin{acknowledgements}
Correspondence and requests for
materials should be addressed to the author (heyl@phas.ubc.ca).
This research has made use of NASA's Astrophysics Data System
Bibliographic Services.
\end{acknowledgements}

\bibliographystyle{apj}
\bibliography{ns,physics,qed,mine}   

\begin{thebibliography}{27}
\expandafter\ifx\csname natexlab\endcsname\relax\def\natexlab#1{#1}\fi

\bibitem[{{den Hartog} {et~al.}(2004){den Hartog}, {Kuiper}, {Hermsen}, \&
  {Vink}}]{2004ATel..293....1D}
{den Hartog}, P.~R., {Kuiper}, L., {Hermsen}, W., \& {Vink}, J. 2004, The
  Astronomer's Telegram, 293, 1

\bibitem[{{Gavriil} {et~al.}(2002){Gavriil}, {Kaspi}, \&
  {Woods}}]{2002Natur.419..142G}
{Gavriil}, F.~P., {Kaspi}, V.~M., \& {Woods}, P.~M. 2002, \nat, 419, 142

\bibitem[{Heyl \& Hernquist(1997{\natexlab{a}})}]{Heyl97hesplit}
Heyl, J.~S. \& Hernquist, L. 1997{\natexlab{a}}, \prd, 55, 2449

\bibitem[{Heyl \& Hernquist(1997{\natexlab{b}})}]{Heyl97index}
---. 1997{\natexlab{b}}, \jpa, 30, 6485

\bibitem[{Heyl \& Hernquist(1998{\natexlab{a}})}]{Heyl97analns}
---. 1998{\natexlab{a}}, \mn, 300, 599

\bibitem[{Heyl \& Hernquist(1998{\natexlab{b}})}]{Heyl98shocks}
---. 1998{\natexlab{b}}, \prd, 58, 043005 (10 pages)

\bibitem[{Heyl \& Hernquist(1999)}]{Heyl98mhd}
---. 1999, \prd, 59, 045005 (5 pages)

\bibitem[{Heyl \& Hernquist(2005{\natexlab{a}})}]{Heyl05sgr}
---. 2005{\natexlab{a}}, \mn, 362, 777

\bibitem[{Heyl \& Hernquist(2005{\natexlab{b}})}]{Heyl03sgr}
---. 2005{\natexlab{b}}, \apj, 618, 463

\bibitem[{Heyl \& Kulkarni(1998)}]{Heyl98decay}
Heyl, J.~S. \& Kulkarni, S.~R. 1998, \apjl, 506, 61

\bibitem[{{Hulleman} {et~al.}(2000){Hulleman}, {van Kerkwijk}, \&
  {Kulkarni}}]{2000Natur.408..689H}
{Hulleman}, F., {van Kerkwijk}, M.~H., \& {Kulkarni}, S.~R. 2000, \nat, 408,
  689

\bibitem[{{Hurley} {et~al.}(1999){Hurley}, {Li}, {Kouveliotou}, {Murakami},
  {Ando}, {Strohmayer}, {van Paradijs}, {Vrba}, {Luginbuhl}, {Yoshida}, \&
  {Smith}}]{1999ApJ...510L.111H}
{Hurley}, K., {Li}, P., {Kouveliotou}, C., {Murakami}, T., {Ando}, M.,
  {Strohmayer}, T., {van Paradijs}, J., {Vrba}, F., {Luginbuhl}, C., {Yoshida},
  A., \& {Smith}, I. 1999, \apjl, 510, L111

\bibitem[{{Hurley} {et~al.}(1996){Hurley}, {Li}, {Vrba}, {Luginbuhl},
  {Hartmann}, {Kouveliotou}, {Meegan}, {Fishman}, {Kulkarni}, {Frail},
  {Bowyer}, \& {Lampton}}]{1996ApJ...463L..13H}
{Hurley}, K., {Li}, P., {Vrba}, F., {Luginbuhl}, C., {Hartmann}, D.,
  {Kouveliotou}, C., {Meegan}, C., {Fishman}, G., {Kulkarni}, S., {Frail}, D.,
  {Bowyer}, S., \& {Lampton}, M. 1996, \apjl, 463, L13+

\bibitem[{{Jung}(1989)}]{1989ApJ...338..972J}
{Jung}, G.~V. 1989, \apj, 338, 972

\bibitem[{{Kaspi} {et~al.}(2003){Kaspi}, {Gavriil}, {Woods}, {Jensen},
  {Roberts}, \& {Chakrabarty}}]{2003ApJ...588L..93K}
{Kaspi}, V.~M., {Gavriil}, F.~P., {Woods}, P.~M., {Jensen}, J.~B., {Roberts},
  M.~S.~E., \& {Chakrabarty}, D. 2003, \apjl, 588, L93

\bibitem[{{Kuiper} {et~al.}(2006){Kuiper}, {Hermsen}, {den Hartog}, \&
  {Collmar}}]{2006ApJ...645..556K}
{Kuiper}, L., {Hermsen}, W., {den Hartog}, P.~R., \& {Collmar}, W. 2006, \apj,
  645, 556

\bibitem[{{Kuiper} {et~al.}(2004){Kuiper}, {Hermsen}, \&
  {Mendez}}]{2004ApJ...613.1173K}
{Kuiper}, L., {Hermsen}, W., \& {Mendez}, M. 2004, \apj, 613, 1173

\bibitem[{{Mereghetti} {et~al.}(2004){Mereghetti}, {Gotz}, {Mirabel}, \&
  {Hurley}}]{2004astro.ph.11695M}
{Mereghetti}, S., {Gotz}, D., {Mirabel}, I.~F., \& {Hurley}, K. 2004, ArXiv
  Astrophysics e-prints, astro-ph/0411695

\bibitem[{{Molkov} {et~al.}(2004){Molkov}, {Hurley}, {Sunyaev}, {Shtykovsky},
  \& {Revnivtsev}}]{2004astro.ph.11696M}
{Molkov}, S., {Hurley}, K., {Sunyaev}, R., {Shtykovsky}, P., \& {Revnivtsev},
  M. 2004, ArXiv Astrophysics e-prints, astro-ph/0411696

\bibitem[{{Murakami} {et~al.}(1994){Murakami}, {Tanaka}, {Kulkarni}, {Ogasaka},
  {Sonobe}, {Ogawara}, {Aoki}, \& {Yoshida}}]{1994Natur.368..127M}
{Murakami}, T., {Tanaka}, Y., {Kulkarni}, S.~R., {Ogasaka}, Y., {Sonobe}, T.,
  {Ogawara}, Y., {Aoki}, T., \& {Yoshida}, A. 1994, \nat, 368, 127

\bibitem[{{{\"O}zel}(2004)}]{2004astro.ph..4144O}
{{\"O}zel}, F. 2004, ArXiv Astrophysics e-prints, astro-ph/0404144

\bibitem[{{Potekhin} {et~al.}(2004){Potekhin}, {Lai}, {Chabrier}, \&
  {Ho}}]{2004ApJ...612.1034P}
{Potekhin}, A.~Y., {Lai}, D., {Chabrier}, G., \& {Ho}, W.~C.~G. 2004, \apj,
  612, 1034

\bibitem[{{Rothschild} {et~al.}(1994){Rothschild}, {Kulkarni}, \&
  {Lingenfelter}}]{1994Natur.368..432R}
{Rothschild}, R.~E., {Kulkarni}, S.~R., \& {Lingenfelter}, R.~E. 1994, \nat,
  368, 432

\bibitem[{{Sanbonmatsu} \& {Helfand}(1992)}]{1992AJ....104.2189S}
{Sanbonmatsu}, K.~Y. \& {Helfand}, D.~J. 1992, \aj, 104, 2189

\bibitem[{Thompson \& Duncan(1995)}]{Thom95}
Thompson, C. \& Duncan, R.~C. 1995, MNRAS, 275, 255

\bibitem[{Thompson \& Duncan(1996)}]{Thom96}
---. 1996, ApJ, 473, 322

\bibitem[{{Usov} \& {Melrose}(1995)}]{1995AuJPh..48..571U}
{Usov}, V.~V. \& {Melrose}, D.~B. 1995, Australian Journal of Physics, 48, 571

\end{thebibliography}

\end{document}